\documentclass[twocolumn]{aastex62}
\usepackage{lineno}
\usepackage{mathtools}
\usepackage{amsmath}
\usepackage{appendix}
\usepackage{comment}

\begin{document}

\title{Understanding how fast black holes spin by analysing data from the second gravitational-wave catalogue}
\author{Charlie Hoy}
\affiliation{Cardiff University, Cardiff CF24 3AA, UK}
\author{Stephen Fairhurst}
\affiliation{Cardiff University, Cardiff CF24 3AA, UK}
\author{Mark Hannam}
\affiliation{Cardiff University, Cardiff CF24 3AA, UK}
\author{Vaibhav Tiwari}
\affiliation{Cardiff University, Cardiff CF24 3AA, UK}
\date{\today}

% abstract
\begin{abstract}\label{sec:abstract}
    The Advanced LIGO and Virgo detectors have now observed approximately 50 black-hole-binary mergers, from which we can begin to infer how rapidly astrophysical black holes spin. The LIGO-Virgo Collaboration (LVC) analysis of detections up to the end of the first half of the third observing run (O3a) appeared to uncover a distribution of spin magnitudes that peaks at $\sim$0.2. This is surprising: is there a black-hole formation mechanism that prefers a particular, non-zero spin magnitude, or could this be the cumulative effect of multiple formation processes? We perform an independent analysis of the most recent gravitational-wave catalogue, and find that (a) the support for the LVC spin-magnitude is tenuous; in particular, adding or removing just one signal from the catalogue can remove the statistical preference for this distribution, and (b) we find potential evidence for two spin sub-populations in the observed black holes; one with extremely low spins and one with larger spin magnitudes. We make the connection that these spin sub-populations could be correlated with the mass of the binary, with more massive binaries preferring larger spin magnitudes, and argue that this may provide evidence for hierarchical mergers in the second gravitational-wave catalogue.
\end{abstract}

% introduction
\section{Introduction}\label{sec:introduction}

During the first and second gravitational-wave (GW) observing runs
(O1 and O2)~\citep{abbott2019gwtc} of the Advanced
LIGO~\citep{TheLIGOScientific:2014jea} and Advanced
Virgo~\citep{acernese2014advanced} GW observatories,
the LIGO scientific and Virgo collaborations announced eleven
GW candidates; ten from merging black-hole binaries (BBHs) ~\citep{Abbott:2016blz, abbott2016gw151226, scientific2017gw170104, abbott2017gw170608, abbott2017gw170814, abbott2019gwtc}
and one from a binary neutron
star coalescence~\citep{abbott2017gw170817}.
Independent groups also reported additional
GW candidates~\citep{nitz20202, venumadhav2020new, zackay2019highly, zackay2019detecting}.
By combining parameter estimates for these BBH observations, first attempts at deciphering the astrophysical distribution of black hole spins were conducted~\citep{Farr:2017uvj, BFarrSpin, Tiwari:2018qch, abbott2019binary, Fairhurst:2019srr, biscoveanu2020new, Garcia-Bellido:2020pwq}.
The limited sample size meant that only weak constraints could be placed
on the distribution of black hole spins, although even with only a handful of signals 
it was shown that high spin magnitudes were strongly disfavoured, and there was some 
evidence for spin mis-alignment in binaries. 

Precise measurements of both the spin magnitudes and their mis-alignment with the binary's orbital angular momentum rely on identifying the presence (or lack) of the General Relativistic phenomenon of spin-induced orbital precession ---
the misalignment of the binary's orbital angular momentum and the spins of
each compact object resulting in characteristic modulations to the GWs amplitude
and phase~\citep{apostolatos1994spin}. A direct measurement of spin-induced
orbital precession would then provide a unique insight into the astrophysical distribution of
black hole spins~\citep[e.g.][]{Gerosa:2018wbw,rodriguez2016illuminating}. During the first half of the third GW observing run (O3a),
a further 39 GW candidates were announced~\citep{abbott2020gwtc2}. However, similar to those from O1 and O2~\citep{LIGOScientific:2018jsj}, most of these detections
remained largely uninformative about the presence of precession~\citep{abbott2020gwtc2}.

\citet{abbott2020gwtc2prop} recently showed, through a hierarchical Bayesian
analysis~\citep{Thrane:2018qnx,Mandel2019,Vitale:2020aaz}, that there is clear evidence that
the population of known BBHs includes misaligned spins, despite no single event unambiguously 
exhibiting evidence for precession. By assuming a population model where the spin magnitude of each
black hole is described by a beta function~\citep{Wysocki:2018mpo} and the orientation by a model
allowing for both isotropic and aligned spins~\citep{Talbot:2017yur}, it was shown that the most likely
spin distribution has a peak in the spin magnitude at $\sim 0.2$, with preference for primarily aligned
spins (although there is non-vanishing support for angles $>90^\circ$ indicating the presence of
misaligned component spins). \citet{Roulet:2021hcu} and \citet{Galaudage:2021rkt} later
challenged this point of view, showing that all binaries observed to date are consistent
with two spin populations: one with negligible black hole spin and a second with
spins preferentially aligned with the orbital angular momentum. \citet{Galaudage:2021rkt}
estimated that 70 -- 90\% of merging binaries contain black holes with negligible spin and
the black holes in the second sub-population have spins $\sim 0.5$ with orientations
preferentially (but not exactly) aligned to the orbital angular momentum.
\citet{Callister:2021fpo} also searched for correlations within the black hole
spin distribution and found that the binaries mass ratio is correlated
with the black hole spin at 98.7\% credibility with more unequal mass binaries
exhibiting systematically larger black hole spin.

In this paper we analyse the results from \citet{abbott2020gwtc2prop} and
link our findings to the conclusions found from other works.
We opt to extend previous
analyses~\citep[see e.g.][]{Farr:2017uvj, Tiwari:2018qch} by including
the effects of spin-induced orbital precession
and perform a detailed model selection analysis for two reasons:
firstly we are able to provide odds ratios to
show by how much one model is preferred to another, and secondly we are able to
``open the black box'' of the results presented in \citet{abbott2020gwtc2prop}
by showing how each GW candidate contributes to the final result.

Since there are still only a limited
number of binary black hole mergers available, which makes inferring the \emph{true} black
hole spin distribution challenging, we select a series of spin models which aim to identify
the core features of the spin distribution. We choose spin models with
isotropic or preferentially aligned spins with several choices for the
distribution of spin magnitudes and compare our results to the
most likely distribution from \cite{abbott2020gwtc2}. We test the robustness of our
results by repeating
the analysis using two parameterisations of precession.

We show that if we use the same set of BBHs, we obtain the same conclusion as \citet{abbott2020gwtc2prop} irrespective of which parameterisation of precession we use, i.e., the current population of binary black holes prefers a spin distribution model with mild preference for aligned spins and spin magnitudes peaking at $\sim 0.2$. However, the preference 
for this distribution disappears if we include
GW190814 in the analysis, and can be disfavoured by as much as 14:1 if
GW190517\_055101 is removed.

Secondly, we show that our results agree with those presented in \citet{Roulet:2021hcu} and \citet{Galaudage:2021rkt} and show that there is evidence that the black hole population is consistent with two
spin sub-populations, which, when combined, gives an overall preference for the distribution in
\citet{abbott2020gwtc2prop}. We highlight that the majority of events ($\sim 80\%$) prefer
a spin distribution with extremely low spin magnitudes while
several show preference for larger spins. We demonstrate that
these two spin sub-populations could be correlated with the mass of the
system with low mass events preferring extremely low spin magnitudes
and high mass events preferring larger spins. Of the models that we use,
the heavier binaries prefer isotropic rather than aligned spin orientations.

This paper is structured as follows: in Section~\ref{sec:method} we
briefly summarize the details behind a model selection analysis and
document which GW candidates and spin distribution models are used
in this study. In Section~\ref{sec:results} we describe our results and provide an understanding into how fast black holes spin. We then
conclude in Section~\ref{sec:conclusion}.

% method
\section{Method}\label{sec:method}

\begin{figure*}
    \centering
    \includegraphics[width=0.47\textwidth]{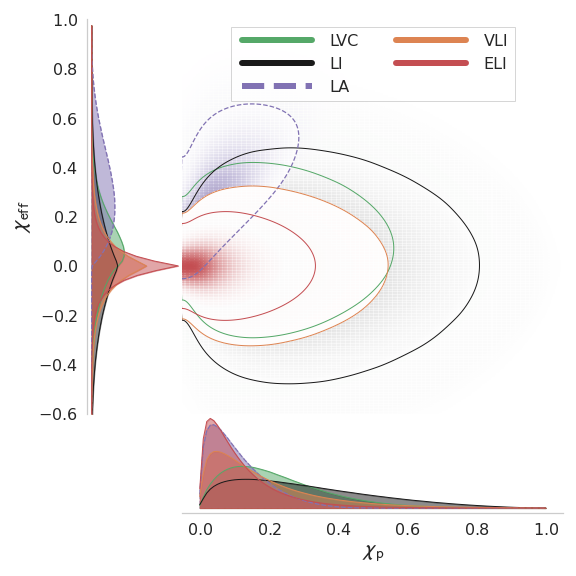}
    \includegraphics[width=0.5\textwidth]{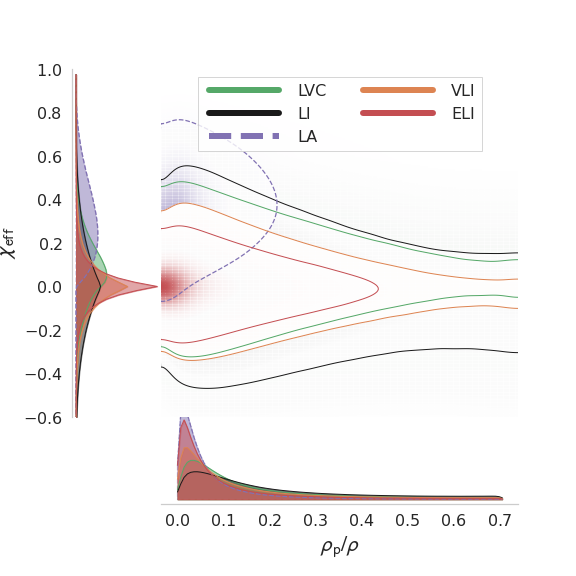}
    \caption{2-dimensional probability density functions (PDFs) showing support across the
    \emph{Left}: $\chi_{\mathrm{p}}$--$\chi_{\mathrm{eff}}$ and \emph{Right}: $\rho_{\mathrm{p}} / \rho$--$\chi_{\mathrm{eff}}$ parameter space for a selection of the different spin distributions used in our analysis -- LVC refers to the mean inferred
    spin distribution from~\citet{abbott2020gwtc2prop}, ELI, VLI and LI are all distributions
    with isotropic spin orientations with either \emph{extremely low}, \emph{very low} or
    \emph{low} spin magnitudes respectively and LA is a \emph{low} spin magnitude
    distribution with spins nearly aligned with the orbital angular momentum (maximum misalignment angles of $30^{\circ}$), see text for definitions. Each contour encloses 90\% of the distribution. We show the
    preferred model from~\citet{abbott2020gwtc2prop} in
    green. Gaussian
    kernel density estimates are used to estimate the probability density. The $\rho_{\mathrm{p}} / \rho$ marginalized
    PDF for the LA spin distribution is truncated to allow for distributions of other models to be seen.
    }
    \label{fig:prior_distribution}
\end{figure*}

We use Bayesian model selection to calculate the odds ratio between
nine different spin distributions using the publicly released
posterior samples from GWTC-2, made available through 
the Gravitational Wave Open Science Center~\citep[GWOSC;][]{vallisneri2015ligo, LIGOScientific:2019lzm}.
As in~\citet{abbott2020gwtc2prop}, we only consider BBHs with
false alarm rates (FARs) $<1 \mathrm{yr}^{-1}$. This means we exclude
2 marginal events included in~\citet{abbott2020gwtc2}:
GW190719\_215514 and GW190909\_114149 (we 
consider GW190426\_152155 to be a marginal neutron star black hole
candidate). Unlike~\citet{abbott2020gwtc2prop} where GW190814 was excluded in their spin distribution analysis, we consider how our results change if GW190814 is included in the population. This is because
GW190814 is most likely (71\%) the result of a BBH merger~\citep{Abbott:2020khf}.
For each GW candidate considered, we randomly draw $10^{4}$ samples from the
`PublicationSamples' dataset to ensure a consistent number of samples across
the population. For all candidates this meant that we used posterior
samples that were generated using waveform models that at least included
precession~\citep{Hannam:2013oca, Ossokine:2020kjp}, and for the majority of candidates this meant that we used posterior samples that were generated using waveform models that also included
subdominant multipole moments~\citep{Khan:2019kot, Ossokine:2020kjp, Varma:2019csw}; see Table VIII in~\citet{abbott2020gwtc2}. Unless otherwise stated, the odds ratio is compared to
the mean distribution displayed in Fig.~10
of~\citet{abbott2020gwtc2prop}, denoted in this work as the ``LVC'' distribution,
and we do not incorporate the calculated uncertainty on the inferred LVC spin distribution.
Given that population studies of BBHs from O1, O2 and
O3a disfavour highly spinning black
holes~\citep{Farr:2017uvj, Tiwari:2018qch, Fairhurst:2019srr, abbott2020gwtc2prop},
we consider only spin distributions with either low (L)~\citep[consistent with][]{Farr:2017uvj},
very low (VL)~\citep[consistent with][]{LIGOScientific:2018jsj} or extremely low (EL) spin
magnitudes,

\begin{equation}\label{eq:spih_mag}
    \begin{split}
        p_{\mathrm{EL}}(a) &\propto e^{-8a}, \\
        p_{\mathrm{VL}}(a) &\propto e^{-5a}, \\
        p_{\mathrm{L}}(a) &\propto (1 - a), \\
    \end{split}
\end{equation}
where $a = |cS/(Gm^{2})|$ is the spin magnitude of a black hole
with mass $m$ and spin angular momentum $S$. We consider two distributions for the tilt angles: 
nearly aligned (A) and isotropic (I). The nearly aligned distribution is triangular in $\cos\theta$,
with a peak at 1 and taking values between $0.85 \leq \cos\theta \leq 1$. The nearly aligned distribution resembles field binaries -- binaries which are formed from
isolated stellar progenitors and are expected to have spins distributed about the orbital angular momentum
with some unknown misalignment angle~\citep[e.g.][]{Kalogera:1999tq, Mandel:2009nx, Gerosa:2018wbw} -- with maximum 
misalignment angles of $30^{\circ}$.
The isotropic distribution is uniform in $\cos\theta$ between $-1$ and $1$ and resembles dynamic
binaries -- binaries which are expected to have randomly orientated spins and formed when two black holes
become gravitationally bound in dense stellar environments~\citep[e.g.][]{rodriguez2016illuminating}.

As it is difficult to constrain the individual black hole spins at typical
signal-to-noise ratios (SNRs)~\citep{purrer2016can}, we use a mass weighted effective spin
$\chi_{\mathrm{eff}}$ to describe the average projection of spins parallel to
the orbital angular momentum~\citep{ajith2011inspiral}, and we use
two competing quantities to describe the projection of spins perpendicular to the orbital angular momentum: $\chi_{\mathrm{p}}$~\citep{Schmidt:2014iyl} and $\rho_{\mathrm{p}}$~\citep{Fairhurst:2019vut}. $\chi_{\mathrm{p}}$ is widely used
for inferring the occurrence of precession in GW
data~\citep[see e.g.][]{abbott2019gwtc,abbott2020gwtc2}, although alternative metrics have also been proposed~\citep[e.g.][]{Gerosa:2020aiw, Thomas:2020uqj}, and ranges
between 0 (no precession of the orbital plane)
and 1 (maximal precesion). $\rho_{\mathrm{p}}$ is the precession
SNR and describes the
contribution to the total SNR of the system that can be attributed to
precession. It has been shown previously that $\rho_{\mathrm{p}}$ is a
useful quantity for inferring the presence of precession in population
studies~\citep{Fairhurst:2019srr}. $\rho_{\mathrm{p}}$ is calculated by decomposing a GW into two non-precessing
harmonics and
isolating the SNR contained in the harmonic orthogonal to the dominant one.
By deconstructing a precessing gravitational-wave in this form, the characteristic
amplitude and phase modulations can be interpreted as the beating of these
harmonics. If $\rho_{\mathrm{p}}$ is small ($\rho_{\mathrm{p}} \lesssim 2$), the amplitude
of the second harmonic is insignificant and any beating of the harmonics is negligible.
For this case, we would observe a GW that looks like the dominant non-precessing harmonic. We choose to perform two analyses, one
describing where the black hole spins are described by
$\chi_{\mathrm{eff}}$ and $\chi_{\mathrm{p}}$ and another with $\chi_{\mathrm{eff}}$ and $\rho_{\mathrm{p}}$. This is because
it is notoriously difficult to accurately extract precession information
from gravitational-wave signals, especially at relatively low SNRs~\citep[see e.g.][]{vecchio2004lisa,Lang:1900bz,berti2005estimating,Vitale:2014mka,abbott2019gwtc, Pratten:2020igi,Green:2020ptm,Kalaghatgi:2020gsq,Krishnendu:2021cyi},
and by performing two separate analyses we can test the
robustness of our conclusions.

Although $\rho_{\mathrm{p}}$ is dependent
on the GW detector network and its sensitivity, if the harmonics are
close to orthogonal, $\rho_{\mathrm{p}}$ can be scaled by the total SNR $\rho$ to provide
a detector invariant quantity $\rho_{\mathrm{p}} / \rho$ (see Eq. 39 in~\citet{Fairhurst:2019vut}, noting that $\rho$ is denoted as $\rho_{2\mathrm{harm}}$).
We therefore choose to parameterise precession in this work by $\rho_{\mathrm{p}} / \rho$ since it allows for results that are independent of
the detector network and the chosen detector sensitivity. This implies that the
$\rho_{\mathrm{p}} \gtrsim 2$ criterion used in previous works for quantifying the measurability of
precession~\cite[see e.g.][]{abbott2020gwtc2,LIGOScientific:2020stgf,Abbott:2020khf,Fairhurst:2019vut,Fairhurst:2019srr,Green:2020ptm}
becomes $\rho_{\mathrm{p}} / \rho \gtrsim 2 / \rho$, which is bounded between
$0 \leq \rho_{\mathrm{p}} / \rho \leq 1/\sqrt{2}$, where the upper bound implies equal power in both
harmonics. Consequently, for systems with large
$\rho_{\mathrm{p}} / \rho$, precession contributes significantly to the total SNR of the system. 

Figure~\ref{fig:prior_distribution} shows how a subset of these spin distributions
vary across the $\chi_{\mathrm{p}}$--$\chi_{\mathrm{eff}}$ and $\rho_{\mathrm{p}}/\rho$--$\chi_{\mathrm{eff}}$ parameter
space. The aligned distributions can easily be distinguished
from the isotropic distributions as $\chi_{\mathrm{eff}}>0$ and
$\rho_{\mathrm{p}}/\rho$ is small by definition.

Following the methodology described in~\citet{Tiwari:2018qch},
we calculate the odds ratio between two spin distribution models
$\lambda_{1}$ and $\lambda_{2}$ as,
\begin{eqnarray}\label{eq:odds_ratio}
    O & = & \frac{p(\lambda_{1}|\{\mathbf{d}\})}{p(\lambda_{2}|\{\mathbf{d}\})} \\
    & \approx & \left[\frac{V_{\mathrm{pop}}(\lambda_{1})}{V_{\mathrm{pop}}(\lambda_{2})}\right]^{-N} \prod_{i=1}^{N}\left[\frac{\sum_{j}p(\theta^{j}_{i}|\lambda_{1})/\pi(\theta_{i}^{j})}{\sum_{j}p(\theta^{j}_{i}|\lambda_{2})/\pi(\theta_{i}^{j})} \right] \times \left[\frac{p(\lambda_{1})}{p(\lambda_{2})}\right], \nonumber
\end{eqnarray}
where $p(\lambda|\{\boldsymbol{d}\})$ is the
posterior distribution for the model $\lambda$ given a
set of BBH observations $\{\boldsymbol{d}\}$, $V_{\mathrm{pop}}(\lambda)$
is the sensitive volume for the model $\lambda$,
$\theta^{j}_{i}$ is the $j\mathrm{th}$ posterior sample for observation
$i$, $\sum_{j}p(\theta^{j}_{i}|\lambda) / \pi(\theta^{j}_{i})$ is the
sum over posterior samples re-weighted from the default \emph{prior universe} used
in the LVC analyses $\pi(\theta^{j}_{i})$ (LAL prior), to the \emph{universe} assuming a given
model $\lambda$ $p(\theta^{j}_{i}|\lambda)$,
$p(\lambda)$ is the prior on the model and we restrict
$\boldsymbol{\theta} = (\mathcal{M}, q, \chi_{\mathrm{eff}},
[\chi_{\mathrm{p}}, \rho_{\mathrm{p}}/\rho], \iota)$.
As with~\citet{Tiwari:2018qch}, we assume all models are equally
likely i.e., $p(\lambda) = 1, \forall \lambda$. In order to evaluate
$p(\theta^{j}_{i}|\lambda)$, we generate a \emph{universe} for each spin
distribution model consisting of $10^{7}$ randomly chosen binaries and compute
a five dimensional Gaussian Kernel Density Estimate (KDE).

For LVC parameter estimation analyses, the prior on $m_{1}$ and
$m_{2}$ is taken to be flat and spin vectors are assumed to be uniform in spin magnitude
and isotropic on the sphere~\citep[see Appendix B.1 of][]{abbott2019gwtc}.
When generating a \emph{universe} for a given
model $\lambda$, we use an identical mass distribution
but vary the spin magnitude and orientation vectors (see
Sec.~\ref{sec:reweight} for an analysis that uses an astrophysical mass
distribution).
All other binary parameters are randomly drawn
from the same distributions as used in~\citet{abbott2019gwtc}. 

The sensitive volume $V_{\mathrm{pop}}(\lambda)$ is essential for accounting
for selection effects. It is estimated numerically by injecting GW signals drawn
from model $\lambda$ into GW strain data and searching for them assuming a
given detection threshold~\citep{Tiwari:2017ndi}.
Currently search pipelines employ non-precessing waveform approximants for
matched filtering~\citep{Usman:2015kfa, Messick:2016aqy}. This means that
current techniques to estimate the sensitive volume omit precession (although
see \citet{Gerosa:2020pgy}, which suggests an alternative method that includes
precession). Since precessing signals will be recovered at lower probabilities
than an equivalent precessing search pipeline, we can expect that
the sensitive volume will be underestimated for systems where precession effects
are observable~\citep{Bustillo:2016gid}. However, \citet{Fairhurst:2019vut} argue
that for signals with
low $\rho_{\mathrm{p}}$ this effect is minimal. Given that for most
models used in this paper $\rho_{p} / \rho$ is small, we approximate
$V_{\mathrm{pop}}(\lambda)$ by $V_{\mathrm{pop}}(\lambda^{\mathrm{np}})$:
the sensitive volume for the non-precessing equivalent $\lambda$.

$V_{\mathrm{pop}}(\lambda^{\mathrm{np}})$ as a function of each spin distribution
model shows the trends we expect: we find that as the spin magnitude increases and
spin orientation becomes more aligned, the sensitive volume increases. This
is expected since binaries with a larger aligned spin (larger $\chi_{\mathrm{eff}}$)
can be observed at a greater distance~\citep{ajith2011inspiral}. Since the ELI spin distribution 
leads to a population with the smallest aligned spin, it has the lowest sensitive volume. 
The odds ratio calculation in Eq.~\ref{eq:odds_ratio} involves dividing by the model's sensitive volume. This means that for cases where the sum over re-weighted posterior samples ($\sum_{j}p(\theta^{j}_{i}|\lambda) / \pi(\theta^{j}_{i})$) is similar between two spin models, the model with the lower sensitive volume is preferred. Although for a single event this effect may be minuscule, for 44 observations the odds ratio may increase substantially, e.g., the smaller sensitive volume for ELI over LVC contributes a factor $\sim 10$ to the Bayes factor calculation.

% results and discussion
\section{Results and Discussion}\label{sec:results}

\begin{figure}[t!]
    \centering
    \includegraphics[width=0.49\textwidth]{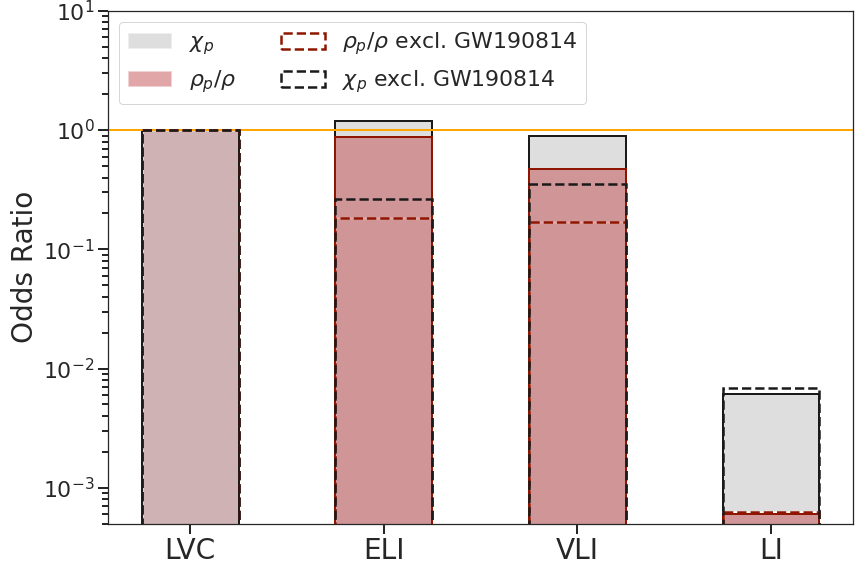}
    \caption{Odds ratios for different spin distributions in reference to the preferred model in~\citet{abbott2020gwtc2prop} (LVC), see text
    for model definitions. The grey and red bars indicate the inferred odds ratios when $\chi_{\mathrm{p}}$ and $\rho_{\mathrm{p}} / \rho$ is used to parameterise precession respectively. Solid lines show the inferred odds ratio when GW190814 is included in the analysis and dashed lines show the inferred odds ratio when GW190814 is excluded. Only the largest four odds ratio are shown. Models that are not shown have odds ratios $<10^{-5}:1$.
    }
    \label{fig:odds}
\end{figure}

\begin{figure}
    \centering
    \includegraphics[width=0.403\textwidth]{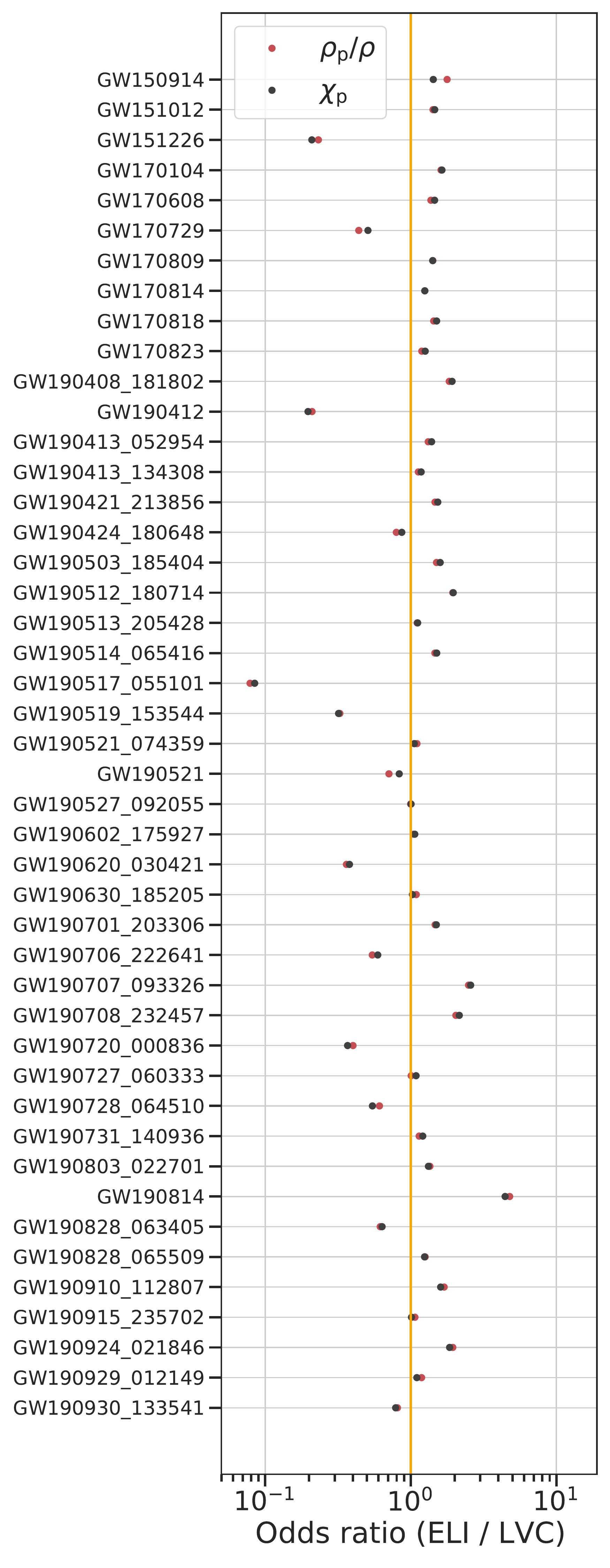}
    \caption{Odds ratios for ELI against
    LVC for each binary black hole candidate considered in this analysis~\citep{abbott2020gwtc2}. An orange line shows an
    odds ratio of $1$ meaning that neither model is preferred. An odds ratio greater than
    $1$ shows preference for ELI over LVC. In both cases odds ratios
    are calculated using two different paramerisations of precession:
    $\chi_{\mathrm{p}}$ (grey) as used in~\citet{abbott2020gwtc2}
    and $\rho_{\mathrm{p}} / \rho$ (red). Odds ratios are calculated using the
    posterior samples released as part of GWTC-2~\citep{abbott2020gwtc2, vallisneri2015ligo, LIGOScientific:2019lzm}.}
    \label{fig:event_by_event}
\end{figure}

Figure~\ref{fig:odds} shows odds ratios with
respect to the LVC spin distribution for two different
metrics for precession: $\chi_{\mathrm{p}}$ and
$\rho_{\mathrm{p}} / \rho$. Models that are not shown have odds ratios $<10^{-5}:1$.  Our analysis shows that if we assume that all
black holes originate from the same population, models with aligned spins are strongly disfavoured ($>10^{7}:1$ in favour of isotropic spins) as well as ``low'' spin magnitudes ($>10^{3}:1$ in favour of ``very low'' or ``extremely low'' magnitudes). Only spin distributions with isotropic spin orientations and very low, or extremely low spin magnitudes are broadly consistent with the observations.  However, the details of which spin distribution
best describes the population depends upon the detailed choice of which events are
included in the analysis. Specifically, when GW190814 is
excluded from our analysis, of the distributions considered, the LVC spin distribution is marginally preferred.   
Our analysis infers that when GW190814 is excluded, both VLI and ELI are disfavoured with odds
ratios $2.8:1$ and $3.8:1$ for the $\chi_{\mathrm{p}}$ analysis and $5.9:1$ and $5.5:1$ for
the $\rho_{\mathrm{p}} / \rho$
analyses respectively. This equates to around a $1 \sigma$ preference for LVC over ELI and $1.4 \sigma$ for VLI.
When GW190814 is included in the population, the extremely low isotropic spin distribution and LVC are equally preferred by the data, with odds ratios ~$\sim 1$ ($1.2:1$ and $0.9:1$ for the $\chi_{\mathrm{p}}$ and $\rho_{\mathrm{p}} / \rho$ analyses respectively) while the very low isotropic model is only slightly dis-favoured.

This is the first main result from our study. The preference for the LVC, extremely low or very low isotropic spin distribution depends sensitively on which signals are included in the analysis. All of these results assume a single population; we will consider multiple sub-populations in Section~\ref{sec:structure}. We now consider these results in more detail.

In Figure~\ref{fig:event_by_event} we show how the odds ratio of ELI vs LVC changes as a function of GW candidate. For the majority of events (32/44) the extremely low spin magnitude distribution is preferred over LVC with 31/44 events having odds ratio between 1 and 3. GW190814 exhibits the strongest preference for extremely low spin magnitudes with an odds ratio $5:1$. On the other hand, there are a handful of events which exhibit a strong preference for larger spins, most notably GW151226, GW190412 and GW190517\_055101.  It is instructive to examine these events in more detail.

GW190517\_055101 has the largest $\chi_{\mathrm{eff}}$ observed so far with $\chi_{\mathrm{eff}}=0.52^{+0.19}_{-0.19}$. This leads to the strongest preference for the LVC distribution among the observed events.  As can be seen in Figure~\ref{fig:prior_distribution} there is significantly greater support for large, positive $\chi_{\mathrm{eff}}$ in the LVC distribution than ELI: the majority of the LVC distribution
supports $\chi_{\mathrm{eff}} > 0$ (70\% compared to 50\% for ELI) and has a
longer tail up to larger $\chi_{\mathrm{eff}}$ ($\sim 0.32$ compared to $\sim 0.1$
for ELI).  If this single event is removed from the analysis, ELI is preferred over LVC with odds ratios
$14:1$ and $11:1$ when GW190814 is included and $3:1$ and $2:1$ when GW190814 is excluded from the analysis for the $\chi_{\mathrm{p}}$ and $\rho_{\mathrm{p}}/\rho$ analyses respectively. Meanwhile, GW190412 supports
$\chi_{\mathrm{eff}} > 0.15$ at 90\% confidence and is consistent with a mildly
precessing system, $\chi_{\mathrm{p}} = 0.31^{+0.19}_{-0.16}$. This event contributes a factor of 5 to the odds ratio in favour of the LVC distribution. GW190412's spin has been discussed at length in previous work~\citep[see e.g.][]{Mandel:2020lhv, Zevin:2020gxf}. Similarly GW151226 has $\chi_{\mathrm{eff}} = 0.18^{+0.20}_{-0.12}$ and again contributes significantly to the preference for larger spins.
Indeed, all candidates that support $\chi_{\mathrm{eff}} > 0$
at more than 90\% probability (GW151226, GW170729, GW190412, GW190517\_055101,
GW190519\_153544, GW190620\_030421, GW170706\_222641,
GW190720\_000836, GW190728\_064510, GW190828\_063405, GW190930\_133541) show a preference for the LVC distribution.

The only event with a large odds ratio in favour of the ELI distribution is GW190814. Uniquely among the events considered, due to the large SNR and the unambiguous identification of higher harmonics, for this binary both the $\chi_{\mathrm{eff}}$ and $\chi_{\mathrm{p}}$ were constrained to be close to zero~\citep{Abbott:2020khf}. As a result it is no surprise that ELI is preferred. In this
particular region of parameter space ELI has $\sim 7\times$ more support than LVC.
Since comparing support in a given region of the parameter space is effectively computing a simplified version of Eq.~\ref{eq:odds_ratio}, we expect this calculation to be indicative of the
odds ratio. It is therefore a good sanity check for our results.

Next we consider the robustness of our results. We see from Figure~\ref{fig:odds} that we obtain
the same conclusions when repeating the analysis using two different metrics of
precession: $\chi_{\mathrm{p}}$ and $\rho_{\mathrm{p}} / \rho$.
As can be seen in Figure~\ref{fig:event_by_event}, the largest difference between these
analyses is for GW190521~\citep{Abbott:2020tfl, Abbott:2020mjq} where the $\chi_{\mathrm{p}}$ and $\rho_{\mathrm{p}} / \rho$ analyses prefer LVC over ELI by $1.2:1$ and $1.4:1$ respectively.
GW190521 is consistent with a merger of two black holes with masses $85^{+21}_{-14}M_{\odot}$ and $66^{+17}_{-18}M_{\odot}$, with effective spins $\chi_{\mathrm{p}}=0.68^{+0.26}_{-0.44}$ and $\chi_{\mathrm{eff}} = 0.08^{+0.27}_{-0.36}$. Owing to the large total mass, GW190521 is very short in duration, with only 4 cycles (2 orbits) within the sensitive frequency band of the GW observatories. In contrast to the $\chi_{\mathrm{p}}$ measurement, the inferred $\rho_{\mathrm{p}} / \rho$ demonstrates a lack of measurable precession, $\rho_{\mathrm{p}} / \rho = 0.16^{+0.33}_{-0.13}$. This is because the short signal implies almost degenerate non-precessing
harmonics (with overlap $0.97^{+0.01}_{-0.03}$) which
leads to a near-zero SNR orthogonal to the dominant harmonic. This difference is explored in
detail in~\citet{Hoy:2021aaa}. Consequently, the
$\chi_{\mathrm{p}}$ analysis infers that GW190521 is consistent with a much larger
spin magnitude distribution than the $\rho_{\mathrm{p}} / \rho$ analysis.

Next we comment briefly on the difference between the $\chi_{\mathrm{p}}$ and $\rho_{\mathrm{p}} / \rho$ analyses on a population level. From Figure~\ref{fig:odds} we see that the difference
in analyses becomes larger for larger spin magnitudes with a $0.4\sigma$, $0.7\sigma$ and $1.7\sigma$
difference between the $\chi_{\mathrm{p}}$ and $\rho_{\mathrm{p}} / \rho$ analyses for the ELI, VLI and LI spin distributions respectively. This is
expected since $\chi_{\mathrm{p}}$ and $\rho_{\mathrm{p}}$ are more likely to
give alternative descriptions regarding the presence of precession in a
gravitational-wave signal at larger rather than lower spin magnitudes,
see e.g. GW190521.

\begin{figure}[t!]
    \centering
    \includegraphics[width=0.48\textwidth]{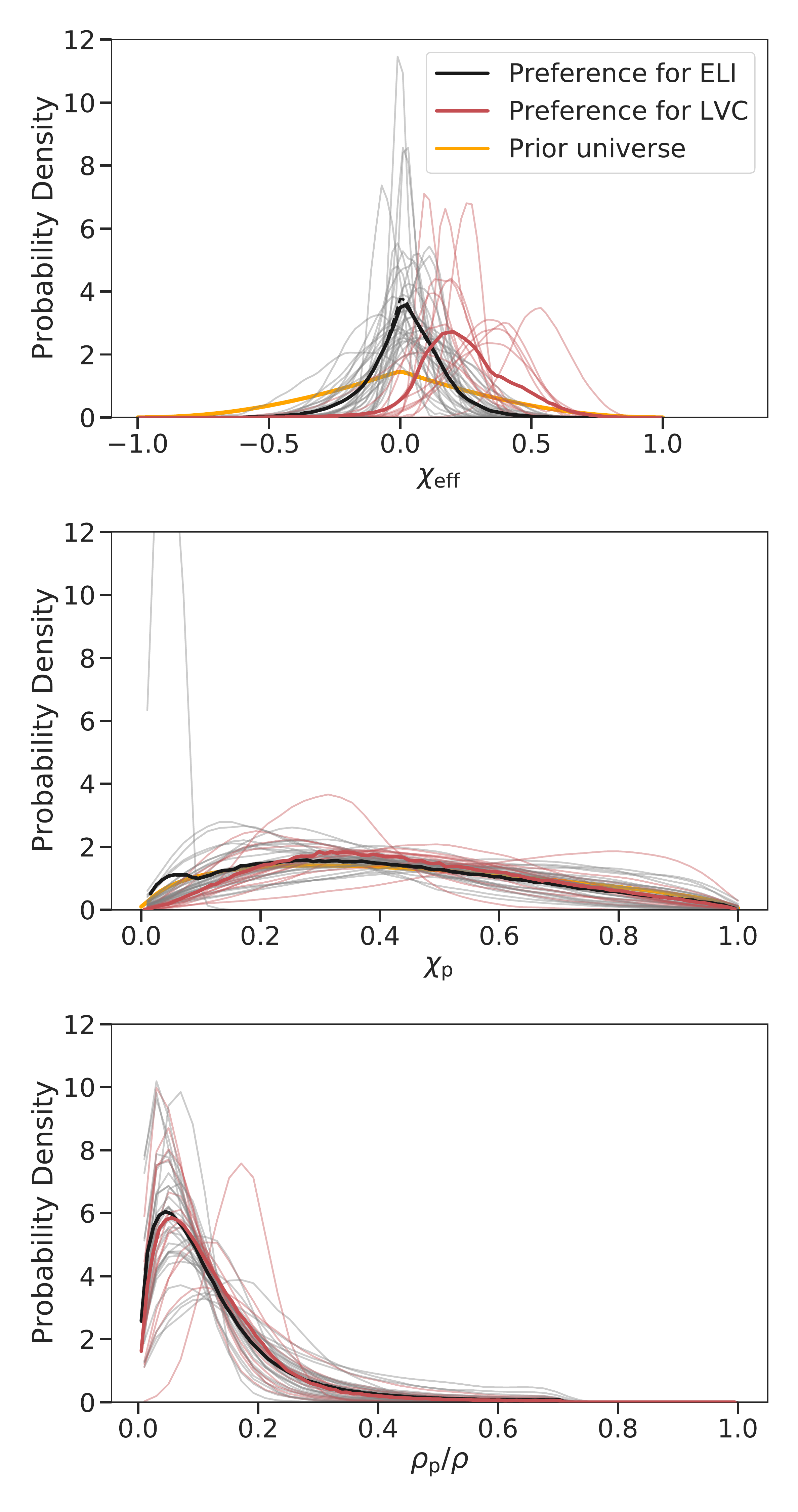}
    \caption{Posterior distributions for \emph{Top}: $\chi_{\mathrm{eff}}$, \emph{Middle}: $\chi_{\mathrm{p}}$, \emph{Bottom}: $\rho_{\mathrm{p}} / \rho$ for binary black hole candidates in the second gravitational-wave catalogue (GWTC-2)~\citep{abbott2020gwtc2, vallisneri2015ligo, LIGOScientific:2019lzm}. Light grey and red traces show the posterior distributions for events which prefer ELI over LVC and LVC over ELI respectively (see Figure~\ref{fig:event_by_event}). Solid black and red curves shows the average of the light grey and red traces respectively. The orange curves show the default $\chi_{\mathrm{eff}}$ and $\chi_{\mathrm{p}}$ priors used in the LVC analyses.
    }
    \label{fig:posteriors}
\end{figure}

\subsection{Reweighting to an astrophysical mass distribution}\label{sec:reweight}

\begin{figure}
    \centering
    \includegraphics[height=0.8104\textheight]{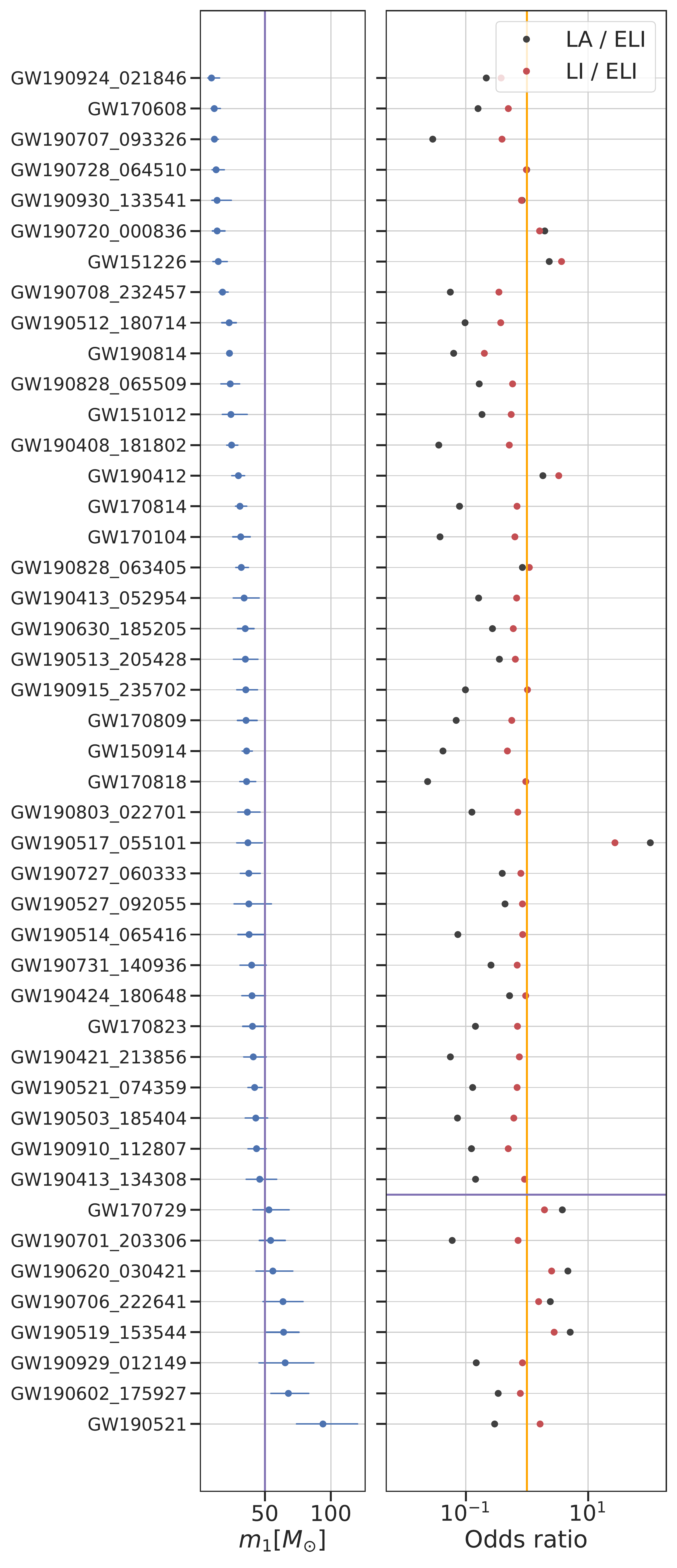}
    \caption{\emph{Left:} A plot showing the mean and 90\% symmetric
    credible intervals for the primary (source frame) mass for all events considered in this analysis. The purple vertical line shows $m_{1} = 50\, M_{\odot}$. \emph{Right}: Odds ratios for LA (black) and LI (red) against
    ELI for each binary black hole candidate considered in this analysis. An orange line shows an
    odds ratio of $1$ meaning that neither model is preferred. An odds ratio less than
    $1$ shows preference for ELI. In both cases the quoted odds ratios
    are an average of the 
    $\chi_{\mathrm{p}}$
    and $\rho_{\mathrm{p}} / \rho$ analyses. Candidates below the purple
    line have primary mass $m_{1} > 50\, M_{\odot}$.}
    \label{fig:event_by_event_sorted_by_mass}
\end{figure}

\begin{figure*}[t!]
    \centering
    \includegraphics[width=0.9\textwidth]{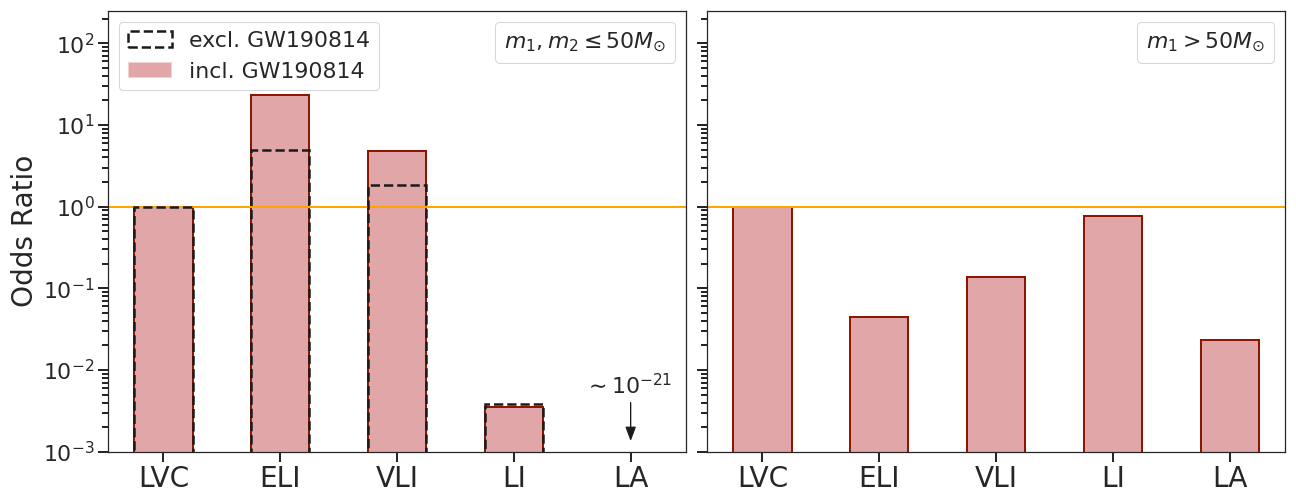}
    \caption{Odds ratios for different spin distributions in reference to the preferred model in~\citet{abbott2020gwtc2prop} (LVC), but considering only BBHs with both component masses less than $50M_{\odot}$ (\emph{Left}) and BBHs with at least one component mass greater than $50M_{\odot}$ (\emph{Right}) based on the arithmetic mean of each BBHs inferred posterior distribution~\citep{abbott2020gwtc2, vallisneri2015ligo, LIGOScientific:2019lzm}.
    Solid red lines show the inferred odds ratio when GW190814 is included in the analysis and dashed black lines show the inferred odds ratio when GW190814 is excluded. For the \emph{Right} panel only the solid red lines are shown since GW190814 has primary mass $m_{1} < 50\, M_{\odot}$.
    The quoted odds ratio's are an average of the $\chi_{\mathrm{p}}$ and $\rho_{\mathrm{p}} / \rho$ analyses.
    }
    \label{fig:high_mass_vs_low_mass}
\end{figure*}

Up until now, all results have used a flat
in $m_{1}$ and $m_{2}$ (with the condition that $m_{1} > m_{2}$) mass distribution. Here, we investigate how the
conclusions vary when the posterior samples are reweighted to an astrophysical
mass distribution. We select a mass distribution where the probability of drawing the
primary mass follows a simple power law $p(m_{1}) \propto m_{1}^{-\alpha}$
and the probability of drawing the secondary mass
is uniform between $5 M_{\odot}$ and $m_{1}$~\citep{Tiwari:2018qch, LIGOScientific:2016dsl}.
For each $\alpha \in [1.5, 2.0, 2.35, 3.0]$  we repeat the analysis above and
identify which spin distribution is preferred.

When reweighting the posterior samples to an astrophysical mass distribution, we expect
to see a preference for lower effective spins. As explained in detail in \citet{Tiwari:2018qch},
this is because close to equal mass ratio binaries are significantly more likely, which, owing to
known degeneracy between the mass ratio and 
aligned-spin~\citep{Cutler:1994ys,Poisson:1995ef,Baird:2012cu}, leads to a preference for
lower effective spins.

As expected, we find that when reweighting to an astrophysical mass prior, the ELI and VLI
spin distributions describe the data well with odds ratios approaching
unity. When GW190814 is included in the analysis, both ELI and VLI are preferred with odds ratios $\sim 4:1$ and $\sim 2:1$ respectively. We also draw similar conclusions to \citet{Tiwari:2018qch} and find that the
result is only mildly dependent on the chosen value of $\alpha$. This
demonstrates that even when reweighting to an astrophysically motivated
mass prior, there is still no strong evidence to prefer one spin model
over another (when GW190814 is excluded from the analysis), and that
the unknown mass distribution of black holes does not cause
a significant effect on the inferred spin distribution.

\subsection{Structure in the preferred spin distribution}\label{sec:structure}

In Figure~\ref{fig:posteriors} we plot the posterior distributions for $\chi_{\mathrm{eff}}$,
$\chi_{\mathrm{p}}$ and $\rho_{\mathrm{p}} / \rho$ for all events used in this analysis. On average the $\chi_{\mathrm{eff}}$ distribution for the events that prefer ELI over LVC (see Figure~\ref{fig:event_by_event}) is strongly peaked at zero with width comparable to ELI (see
Figure~\ref{fig:prior_distribution}). Meanwhile, the average $\chi_{\mathrm{eff}}$ distribution for the events that prefer LVC over ELI
peaks at $\sim 0.2$ with little support for
$\chi_{\mathrm{eff}} \leq 0$. On average there is
no information from precession for binaries that prefer ELI over
LVC or binaries that prefer LVC over ELI with the average
$\chi_{\mathrm{p}}$ posterior resembling the prior and near zero
$\rho_{\mathrm{p}} / \rho$.
This hints at possible sub-populations in the preferred spin distribution: one with
extremely low spins (EL) and one with larger spins (L). This is the same
conclusion found by \citet{Roulet:2021hcu} and \citet{Galaudage:2021rkt}. We therefore investigate these
possible spin sub-populations by calculating odds ratios
between models with extremely low spin (ELI) and models with
larger spins both nearly aligned with the orbital angular
momentum (LA) and spins isotropically distributed (LI). 

We show in Figure~\ref{fig:event_by_event_sorted_by_mass} that
most events in our population prefer a distribution with EL
spins ($\sim 80\%$) while several prefer larger spin magnitudes. Of those events that prefer larger spins,
the nearly aligned distribution (LA) is preferred to the isotropic (LI) with an odds ratio
of $9:1$. 
This result is consistent with the conclusion from
\citet{Galaudage:2021rkt} where it is inferred that a) 70 -- 90\% of 
merging black hole binaries contain black holes with negligible spins and
b) the high spin sub-population has spins preferentially aligned
with the orbital angular momentum. However, the preferentially
aligned spin conclusion is primarily driven by GW190517\_055101
with the odds ratio reducing from $9:1$ to $2:1$ when GW190517\_055101
is excluded from the population.

Interestingly, we see a positive correlation between the binaries
primary mass and the preferred black hole spin distribution with
most low mass binaries preferring distributions with EL
spins and most high mass binaries preferring larger spin magnitudes.
This suggests that the
sub-populations found in \citet{Roulet:2021hcu} and
\citet{Galaudage:2021rkt} could be correlated with the primary mass of the
binary.
In fact, we find that most binaries with primary mass less than
$50 M_{\odot}$ tend to prefer EL spin magnitudes ($33/37$) while most binaries with
primary mass greater than $50 M_{\odot}$ tend to prefer larger spins ($5/8$), see Figure~\ref{fig:high_mass_vs_low_mass}. 
This result is in agreement with the conclusions found by 
\citet{2021ApJ...913L..19T}, which hinted at a possible correlation between
the aligned spin magnitude and the chirp mass of the binary since all events in this high mass sub-population
have chirp mass greater than $32\, M_{\odot}$ -- the point at which the
aligned spin magnitude starts to increase (see Figure 2 of
\citet{2021ApJ...913L..19T}). This positive correlation may suggest evidence for hierarchical
mergers in GWTC-2~\citep{Kimball:2020opk, Kimball:2020qyd, 2021ApJ...913L..19T, Gerosa:2020bjb, Abbott:2020mjq, Fishbach:2017dwv, Doctor:2019ruh}, where the remnant of a previous ``first generation'' binary becomes part of a new one~\citep[e.g.][]{Antonini:2016gqe}; although see~\citet{Jaraba:2021ces} for an alternative mechanism which allows for heavier black holes to have larger spins owing to close hyperbolic encounters spinning up black holes in dense clusters. This is because hierarchical mergers are expected to have a) larger black hole mass and b) larger spins since the remnant of a first generation binary is expected to have mass nearly equal to the sum of its components and spin $a \approx 0.7$ (inherited from the orbital angular momentum of the previous
binary)~\citep{Buonanno:2007sv}. Similarly it is expected that
merging black hole binaries with black hole mass
$m \gtrsim 50 M_{\odot}$ can only be formed through hierarchical mergers since
pair-instability supernova theory~\cite[see e.g.][]{Woosley:2016hmi, Belczynski:2016jno, Gerosa:2017kvu, Abbott:2020tfl, Abbott:2020mjq} prohibits
black holes forming from direct stellar collapse with masses within the range $\sim 50-120 M_\odot$.

From Figure~\ref{fig:high_mass_vs_low_mass} we see that
even when GW190814 is excluded from the analysis, low mass binaries prefer
distributions with EL spins and high mass binaries prefer distribution with larger spin magnitudes. In fact we calculate
that for low mass binaries, EL spin magnitudes are preferred to low
spin magnitudes by $\sim 10^{3}:1$ while for high mass binaries, low spin magnitudes
are preferred to EL spin magnitudes by $\sim 20:1$. Although for high mass binaries there is significantly larger support for 
aligned spins than for low mass binaries, the high mass sub-population prefers
more isotropic spin orientations. This is primarily driven by
GW190701\_203306 and GW190929\_012149 since they both have support for negative $\chi_{\mathrm{eff}}$, $\chi_{\mathrm{eff}} = -0.07^{+0.23}_{-0.29}$ and $\chi_{\mathrm{eff}} = 0.01^{+0.34}_{-0.33}$ respectively, a region of parameter space which is
not permitted by the aligned spin models used in this analysis (see Figure~\ref{fig:prior_distribution}). Since the high
mass sub-population prefers isotropic spin distributions this suggests that
if originated from hierarchical mergers, they are likely to have formed in dense
stellar clusters, where the spins are predicted to be isotropic, rather
than in the accretion disks surrounding Active Galactic Nuclei where the spins are predicted to be aligned with the orbital angular momentum~\citep{Wang:2021clu}.

\citet{abbott2020gwtc2prop} also investigated whether there is evidence for a mass dependence in the BH spin distribution through a hierarchical Bayesian analysis of the population of known BBHs. Similar to the work presented here, \citet{abbott2020gwtc2prop} also found a preference for higher spin magnitudes in higher mass events, although weaker than what we find in this work (see e.g. Figure 13 of \citet{abbott2020gwtc2prop}). However, since the uncertainty on their measurement was broad, a mass dependence could not be confidently claimed.

We note that this potential mass dependence could arise from systematic errors since higher mass systems have far fewer cycles in the sensitive frequency band of the LIGO--Virgo detectors, making it significantly harder to accurately infer the black hole spin~\citep[see e.g.][]{Abbott:2020mjq}. For example, \citet{Haster:2015cnn} recently found that binaries with a larger total mass tend to infer a larger positive black hole spin. We propose that in order to test this possible correlation, this analysis should be repeated on a simulated population in which all systems, high- and low-mass, have very small spins. If the correlation between mass and spin is not found, our conclusion is not an systematic artifact. We leave this for future work.

\citet{Callister:2021fpo} recently identified an anti-correlation between
the binaries mass ratio and black hole spin where more unequal mass binaries
exhibit a systematically larger $\chi_{\mathrm{eff}}$ at 98.7\% confidence.
Our analysis draws similar conclusions,
and finds evidence for a correlation between the binaries mass ratio and black
hole spin. We find that more unequal mass binaries prefer larger spin magnitudes
than more equal mass binaries. For instance, for binaries with component masses
more unequal than GW190527\_092055 (based on their median value),
8/15 prefer EL spins while 7/15 prefer larger spin magnitudes\footnote{GW190814, GW190929\_012149, GW190828\_065509, GW190513\_205428, GW190924\_021846, GW190512\_180714, GW151012, GW190930\_133541 prefer EL spins while GW190412, GW151226, GW190720\_000836, GW190706\_222641, GW190519\_153544, GW190620\_030421, GW170729 prefer larger spin magnitudes.}
This results in a preference for low spin magnitudes over EL spins by
$2:1$ ($8:1$ if GW190814 is excluded from the analysis). For binaries
with more equal masses than GW190527\_092055, EL spins are preferred for all binaries except for
GW190517\_055101. For this subset of binaries the preference for EL spins is
significant -- $10^{3}:1$. Of those unequal mass binaries with a preference for
larger spin magnitudes, Figure~\ref{fig:event_by_event_sorted_by_mass} shows that
several have primary mass $m_{1} > 50 M_{\odot}$. It is possible that
the apparent anti-correlation between the binaries mass ratio and black hole spin
is a consequence of the observed mass dependence described
above, although unlikely on theoretical grounds since hierarchical mergers are predicted to
generate $\chi_{\mathrm{eff}}$ distributions centered at 0 (see e.g. Figure
7 of \citet{Rodriguez:2019huv}). We note that there is also the possibility that our conclusion is a model dependent effect because there exists an inherent correlation between the $\chi_{\mathrm{eff}}$ and mass ratio of the binary for our models. For example, because our models assume uncorrelated spins, it is harder to produce $\chi_{\mathrm{eff}} = 0$ if the mass ratio is equal than if it is extreme. This is a different effect from the known mass ratio and aligned-spin degeneracy in the gravitational wave likelihood~\citep{Cutler:1994ys,Poisson:1995ef,Baird:2012cu} and we leave an investigation into this effect for future work.

% conclusion
\section{Conclusion} \label{sec:conclusion}

In this work we performed an independent analysis and recomputed
the black hole spin distribution using data from the second gravitational-wave
catalog. We demonstrated that the surprising spin magnitude distribution obtained
from \citet{abbott2020gwtc2prop} is unlikely to be robust since
the inclusion of GW190814 or exclusion of GW190517\_055101 changes the preferred
spin distribution to one with extremely low spins. We then demonstrated that our results
are consistent with those from \citet{Roulet:2021hcu, Galaudage:2021rkt} and
established that there is potential evidence for two spin sub-populations in the observed black holes
--- one with extremely low spins and one with larger spin magnitudes.
We then made the argument that these spin sub-populations could be correlated with the primary
mass of the binary, where we see an increase in spin magnitude for
systems with higher masses, and argued that this may provide evidence for hierarchical mergers in
GWTC-2.  Unlike recent works where hierarchical Bayesian inference has been used to
infer the spin distribution of black holes, we chose to perform a detailed model
selection analysis. We suggest that since there are still only a limited number
of binary black hole observations, a much deeper understanding of the inferred
black hole spin distribution can be achieved with this far simpler approach where it is clear how each GW candidate contributes to the final result. With the fourth gravitational-wave
observing run anticipated to provide a plethora of additional binary
black hole observations, with hopefully many more discoveries at high mass, we may soon be able to scrutinize the potential mass--spin correlation as well as deciphering the underlying black hole spin distribution.

% acknowledgements
\section{Acknowledgements}\label{sec:acknowledgements}

We are grateful to Davide Gerosa and Bernard Schutz for discussions during C. Hoy's Ph.D. defence where this work was first presented. We are also thankful to Fabio Antonini, Rhys Green and Cameron Mills for useful discussions, Chase Kimball for comments on this manuscript and Bernard Schutz for the proposal to test the possible mass-spin correlation. This work was supported by Science and Technology Facilities Council (STFC) grant ST/N005430/1 and European Research Council (ERC)
Consolidator Grant 647839, and we are grateful for computational
resources provided by Cardiff University and LIGO Laboratory and supported by STFC grant ST/N000064/1 and National Science Foundation Grants PHY-0757058 and PHY-0823459.

This research has made use of data, software and/or web tools
obtained from the Gravitational Wave Open Science Center
(https://www.gw-openscience.org/ ), a service of LIGO Laboratory,
the LIGO Scientific Collaboration and the Virgo Collaboration. LIGO
Laboratory and Advanced LIGO are funded by the United States National
Science Foundation (NSF) as well as the Science and Technology
Facilities Council (STFC) of the United Kingdom, the
Max-Planck-Society (MPS), and the State of Niedersachsen/Germany for
support of the construction of Advanced LIGO and construction and
operation of the GEO600 detector. Additional support for Advanced
LIGO was provided by the Australian Research Council. Virgo is
funded, through the European Gravitational Observatory (EGO), by the
French Centre National de Recherche Scientifique (CNRS), the Italian
Istituto Nazionale della Fisica Nucleare (INFN) and the Dutch Nikhef,
with contributions by institutions from Belgium, Germany, Greece,
Hungary, Ireland, Japan, Monaco, Poland, Portugal, Spain.

Plots were prepared with the
Matplotlib~\citep{2007CSE.....9...90H} and PESummary~\citep{Hoy:2020vys} and both {\sc{NumPy}}~\citep{numpy} and {\sc{Scipy}}~\citep{mckinney-proc-scipy-2010} were used in the analysis.

\bibliography{references}

\end{document}